\documentclass[letterpaper]{jpconf}
\usepackage{graphicx}

\begin{document}

\title{Benefits of joint LIGO -- Virgo coincidence searches for burst
and inspiral signals}

\author{F.~Beauville$^5$, 
M.-A.~Bizouard$^7$, 
L.~Blackburn$^3$, 
L.~Bosi$^8$, 
P.~Brady$^4$, 
L.~Brocco$^9$, 
D.~Brown$^{2,4}$,
D.~Buskulic$^5$, 
F.~Cavalier$^7$,
S.~Chatterji$^2$, 
N.~Christensen$^1$, 
A.-C.~Clapson$^7$, 
S.~Fairhurst$^4$, 
D.~Grosjean$^5$, 
G.~Guidi$^6$, 
P.~Hello$^7$, 
E.~Katsavounidis$^3$, 
M.~Knight$^1$, 
A.~Lazzarini$^2$,
N.~Leroy$^7$,
F.~Marion$^5$, 
B.~Mours$^5$, 
F.~Ricci$^9$, 
A.~Vicer\'e$^6$, 
M.~Zanolin$^3$ \\[.3cm]
\begin{center}The joint LIGO/Virgo working group\end{center}}
\vspace{.3cm}
\address{$^1$ Carleton College, Northfield MN 55057 USA}
\address{$^2$ LIGO-California Institute of Technology, Pasadena CA 91125 USA}
\address{$^3$ LIGO-Massachusetts Institute of Technology, Cambridge,
Massachusetts 02139 USA}
\address{$^4$ University of Wisconsin - Milwaukee, Milwaukee WI 53201 USA}
\address{$^5$ Laboratoire d'Annecy-le-Vieux de physique des particules, 
Chemin de Bellevue, BP 110, 74941 Annecy-le-Vieux Cedex France }
\address{$^6$ INFN - Sezione Firenze/Urbino Via G.Sansone 1, I-50019 Sesto
Fiorentino; and/or Universit\`a di Firenze, Largo E.Fermi 2, I-50125 Firenze
and/or Universit\`a di Urbino, Via S.Chiara 27, I-61029 Urbino Italia}
\address{$^7$ Laboratoire de l'Acc\'el\'erateur Lin\'eaire (LAL),
IN2P3/CNRS-Universit\'e de Paris-Sud, B.P. 34, 91898 Orsay Cedex France}
\address{$^8$ INFN Sezione di Perugia and/or  Universit\`a di Perugia, Via A.
Pascoli, I-06123 Perugia Italia}
\address{$^9$ INFN, Sezione di Roma  and/or Universit\`a ``La Sapienza",  P.le
A. Moro 2, I-00185, Roma Italia}


\begin{abstract}

We examine the benefits of performing a joint LIGO--Virgo search for
transient signals.  We do this by adding burst and inspiral signals to
24 hours of simulated detector data.  We find significant advantages to
performing a joint coincidence analysis, above either a LIGO only or
Virgo only search.  These include an increased detection efficiency, at
a fixed false alarm rate, to both burst and inspiral events and an
ability to reconstruct the sky location of a signal.  

\end{abstract}

\section{Introduction}\label{sec:intro}

The first generation of gravitational wave interferometric detectors are
approaching their design sensitivities \cite{ref:ligo, ref:virgo,
ref:geo, ref:tama}.  Once fully commissioned, they will provide
unprecedented sensitivity to gravitational waves in the frequency range
between $10$ and $10,000$ Hz.   The goal of these interferometers is to
make the first direct detection of gravitational wave signals.  It has
long been acknowledged that the chances of detection are increased by
making optimal use of data from all available detectors.  However, there
are many issues to be addressed before this is possible.  First, we must
resolve various technical issues associated with analyzing data from
several detectors with different sensitivities, hardware configurations
and sampling rates.  This has been addressed in joint searches of
LIGO--TAMA coincident data \cite{ref:lt_burst, ref:lt_inspiral}, and
previous comparisons of the LIGO and Virgo burst and inspiral search
pipelines \cite{ref:lv_burst, ref:lv_inspiral}.  In addition, we must
understand how to `optimally' combine the data and results from several
different detectors.  In this paper, we perform a study using simulated
data to compare different strategies of combining results from searches
of LIGO and Virgo data.   

There are numerous advantages to performing a joint search.  First, by
requiring a signal to be observed in several detectors in coincidence we
can work at a far lower false alarm rate than is practical using a
single detector.  Futhermore, if we make use of several detectors, it is
possible to recover the sky location and polarization of the signal.  We
can increase the amount of data available for analysis by performing a
search whenever at least two detectors (in a network of three or more)
are operational.  Also, due to the different alignments of the
detectors, their sensitivity to different parts of the sky varies.
Thus, by requiring a signal to be observed in a subset of the detectors,
we can improve our sky coverage.  Finally, a signal seen in several
widely separated detectors making use of different hardware and analysis
algorithms decreases the chance of it being due to any systematic error
or bias.  As is apparent from above, some of the possible benefits of a
multi-detector search are mutually exclusive.  If we require a
coincidence in all available detectors, we will have the lowest possible
false alarm rate, but at the same time we will actually decrease our
sensitivity as any signal which is poorly aligned for one of the
detectors will be missed in coincidence.  Some of the advantages listed
above arise from using the `and' of all available detectors, while
others come from the `or' combination.  Obviously, we must find a
balance between these two competing regimes.

\begin{figure}
\begin{minipage}{18pc}
\includegraphics[width=18pc]{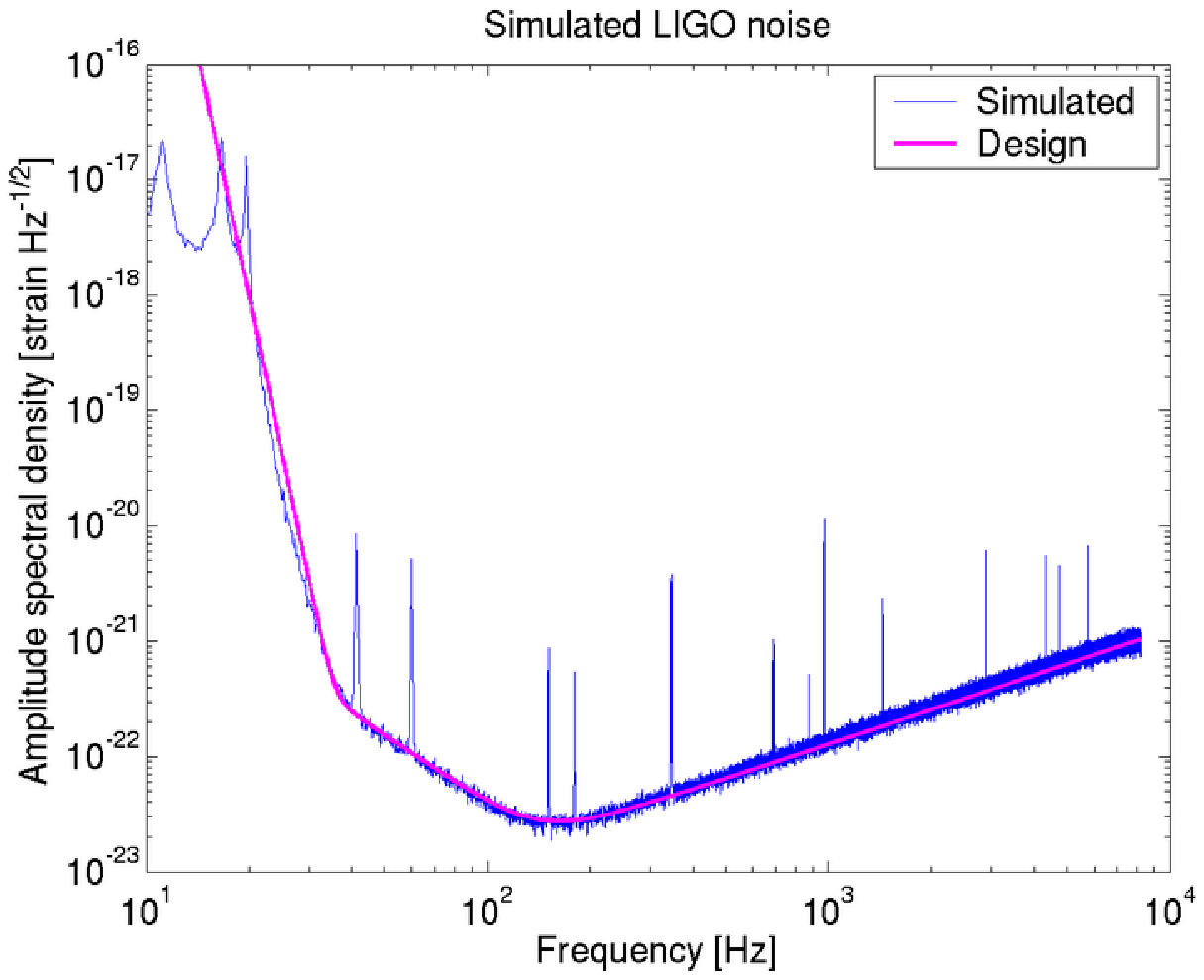}
\caption{The LIGO design sensitivity curve and the spectrum of the
simulated data.}
\label{fig:ligospec}
\end{minipage}\hspace{2pc}
\begin{minipage}{18pc}
\includegraphics[width=18pc]{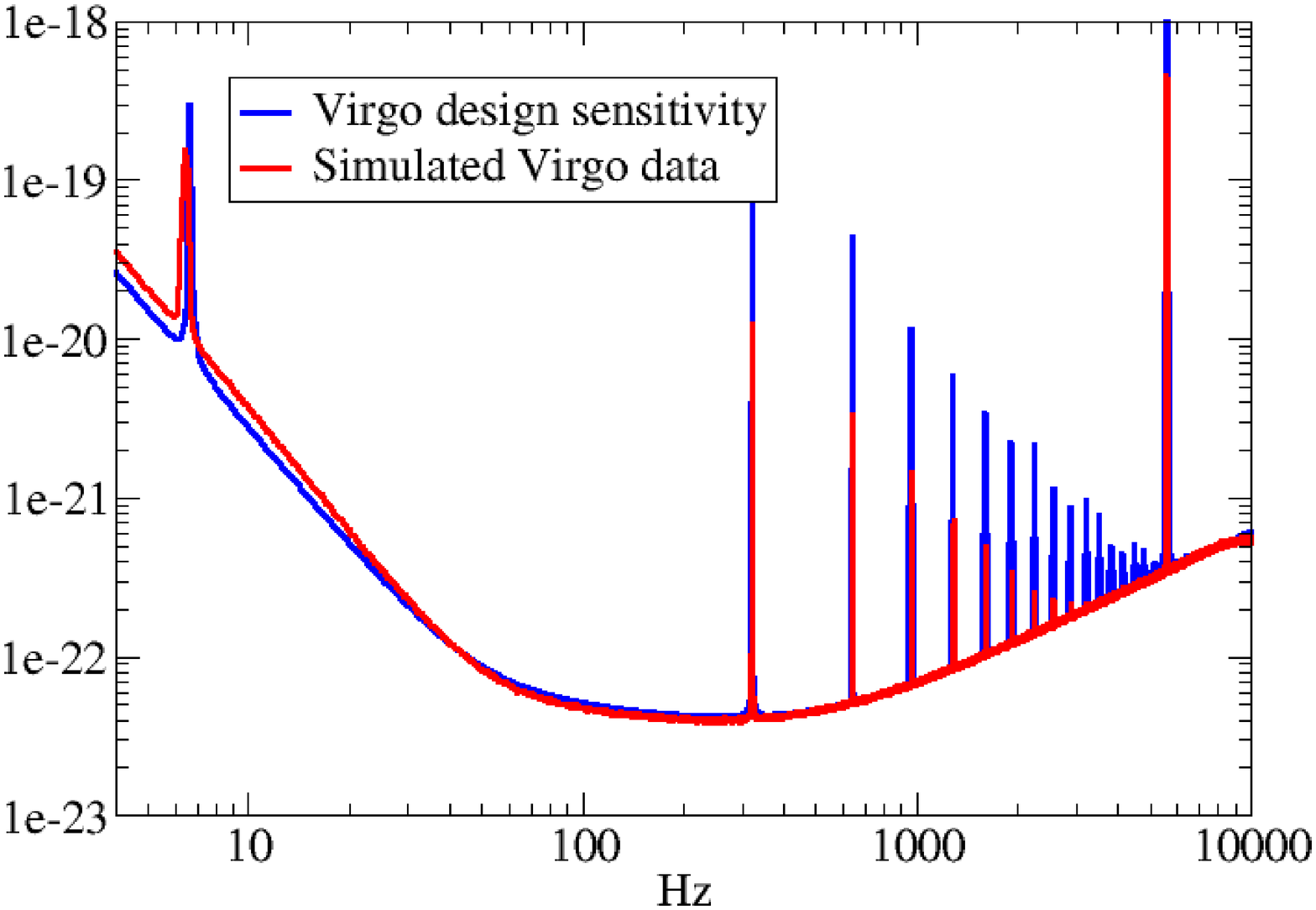}
\caption{The Virgo design sensitivity curve and the spectrum of the
simulated data.}
\label{fig:virgospec}
\end{minipage}
\end{figure}

In this paper, we address the question of how to best combine the
available data from the LIGO and Virgo interferometers.  We explore this
using 24 hours of simulated data for three detectors:  the Virgo
detector (V1) and the two 4 km LIGO detectors, one at Livingtson (L1)
and the other at Hanford (H1).  The noise spectra and design
sensitivities of the LIGO and Virgo detectors are shown in Figures
\ref{fig:ligospec} and \ref{fig:virgospec} respectively.  Into these
data, we inject gravitational wave signals from a variety of burst and
inspiral signals and compare different `and'/`or' combinations of
searches of the three detectors.  

In the discussion above, and throughout the paper, we focus only on a
coincidence analysis.  More specifically, data from each of the
detectors is analyzed independently for candidate events.  Subsequently,
the candidates from each of the single detectors are searched for
coincidences in time (and possibly other parameters).  There are other
approaches which involve a coherent combination of the interferometers'
data streams.  These coherent analyses tend to be computationally costly,
and consequently their use may well be restricted to follow-up analyses
of candidates found in an event-based coincidence search.  The
implementation and testing of coherent search algorithms is a current
research priority, and will be addressed in future publications.

\section{Burst}

Burst search algorithms are designed to identify short duration,
unmodelled gravitational wave burst signals in the detector's data
stream.  There are many different methods of searching for such
unmodelled bursts in the data.  Several of these have been independently
implemented by the LIGO and Virgo collaborations, and a first comparison
of the various methods was made in \cite{ref:lv_burst}.  In this paper,
we will extend that work by examining various methods of combining
results from independent burst searches on the data from the three
detectors H1, L1 and V1.  In particular, we will focus on the benefits
of a multi-detector and multi-site coincidence search, including the
ability to reconstruct the sky location of a signal observed in all
three detectors.

\subsection{Injections}
\label{sec:burst_inj}

\begin{figure}
\includegraphics[width=18pc]{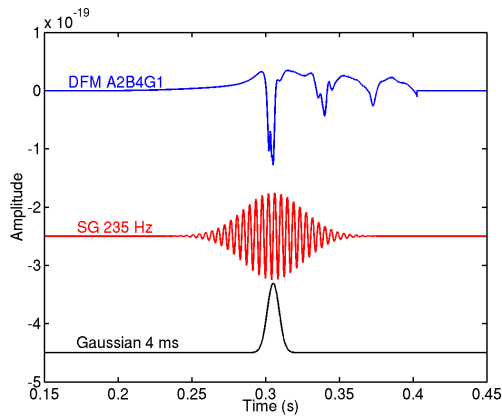}
\begin{minipage}[b]{18pc}
\caption{The burst waveform families used in this analysis.}
\label{fig:burstinj}
\end{minipage}
\end{figure}

In order to test the efficiency of various search methods, as well as
the benefits of a coincidence analysis, it is necessary to add burst
signals to the simulated data streams of the detectors.  In this study,
we inject six different burst signals into the data.  These consist of
two Sine Gaussian signals, one at a frequency of 235 Hz and $Q=5$, the
second with a frequency of 820 Hz and $Q=15$; two Gaussian signals with
widths of 1 and 4 milliseconds; and two supernova core collapse
waveforms, A1B2G1 and A2B4G1, from the catalog of Dimmelmeier, Font and
Mueller \cite{ref:dfm}. The waveform families are illustrated in Figure
\ref{fig:burstinj}.  By using a broad set of waveforms for injection, we
hope to obtain a coarse coverage of the space of possible astrophysical
waveforms.  

We perform burst injections from the direction of the galactic center.
The injections are linearly polarized with uniformly distributed
polarization angle.  We must also specify the amplitude of the
waveforms.  However, these burst waveforms, with the exception of the
supernova core collapse simulation, cannot be normalized to a specific
astrophysical distance.  Instead, we choose a normalization for each
waveform derived from the detectors' sensitivities.  The response of an
interferometric detector to a gravitational wave depends upon the sky
location and polarization of the signal.  Thus, signals from the
galactic center with the same intrinsic magnitude will appear in the
data stream of the detector with different amplitudes, which are
dependent on polarization and (time dependent) sky location of the
source.  We fix the intrinsic amplitude of each waveform by requiring
that there is exactly one injection during the 24 hour data sample with
a signal to noise ratio (SNR) of 10 or greater in all three detectors.
For the supernova core collapse sources, this normalization corresponds
to distances of 4.8 kpc and 3.6 kpc for the A1B2G1 and A2B4G1
simulations respectively.  

\begin{figure}
\includegraphics[width=18pc]{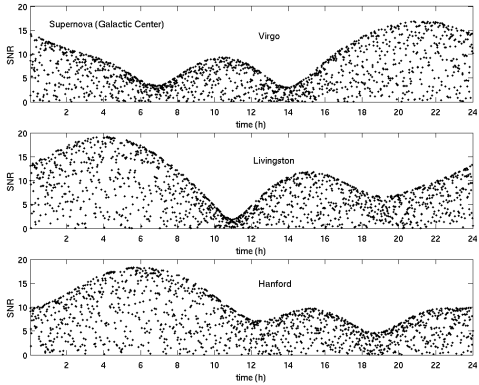}
\begin{minipage}[b]{18pc}
\caption{The daily variation of the signal to noise ratio of injected
supernova signals A2B4G1 from the direction of the galactic center in
the three detectors.  The variation of the maximum SNR is due to the
detector's time varying response to the galactic center, while the
spread (at a given time) is due to the different, random polarizations
of the injected waveforms.} 
\label{fig:dailyvariation}
\end{minipage}
\end{figure}

The signal to noise ratio of the injections in each of the three
detectors is shown in Figure \ref{fig:dailyvariation}.  The detectors'
varying sensitivity to galactocentric sources over the course of the day
modulates the SNR of the simulated singals.  The LIGO detectors at
Livingston and Hanford were designed to have similar orientations,
although they cannot be identical since the detectors are separated by
3000 km.  Their similar orientations give similar directional responses
to gravitational radiation.  Consequently, the SNR distributions of the
injections in H1 and L1 over the course of the day are similar, but by
no means identical.  In particular, both detectors suffer a decrease in
sensitivity to sources from the galactic center at around 11 and 19
hours.  The sensitivity of the Virgo detector to signals from the
galactic center is very different from the LIGO detectors.  For example,
it has a peak in sensitivity at 11 hours when the LIGO detectors are
less sensitive.

\subsection{Single interferometer analysis}
\label{sec:burst_single}

Broadly speaking, burst search algorithms can be characterized as time
domain searches, time/frequency domain searches or correlators.  In this
comparison, we use seven different search algorithms distributed between
these three classes.  We use two time domain methods, the Mean Filter
(MF) and Alternative Linear Filter (ALF).  These identify times at which
the character of the detector data changes.  The time-frequency methods,
PowerFilter and Q-transform, identify areas in the time-frequency plane
with excess power.  Finally, correlators match filter the data using a
specific family of waveforms.  The Peak Correlator (PC) uses gaussian
templates, the Exponential Gaussian Correlator (EGC) uses sine
gaussians, and the Frequency Domain Adaptive Wiener Filter (FDAWF)
algorithm uses Gaussian, zero phase templates.  Details of these methods
and additional references are given in Ref.~\cite{ref:lv_burst}. 

We use all of the methods described above to search for the six
different sets of injected waveforms described in Section
\ref{sec:burst_inj}.  For each algorithm and each waveform, we calculate
the detection efficiency, which is the percentage of injected signals
successfully detected.   To make the results comparable, all searches
are performed with a fixed single-detector false alarm threshold of 0.1
Hz.  The results from the different injected waveforms are comparable,
although different search algorithms are better suited to detecting, and
consequently are more sensitive to, different injected waveforms.  A
full comparison of the results from different search algorithms and
injected waveforms will be presented in a future publication
\cite{ref:lv_burst_coh}.  To simplify the presentation in this paper we
will restrict our attention to one waveform, the supernova core collapse
waveform A2B4G1. For the injected population shown in Figure
\ref{fig:dailyvariation}, the search efficiencies for the three
detectors are given in Table \ref{tab:burst_single_eff}.  In the table,
we give the maximum efficiency obtained by one of the algorithms, as
well as the average of all the search algorithms used.  The best
efficiency for the three detectors is similar, at around 60\%.
Additionally, the average efficiency is only a few percent lower than
the best, showing that the performance difference between various search
algorithms is not too significant.

\begin{table}[ht]
\begin{center}
\begin{tabular}{c c c c}
\hline \hline
 & H1 & L1 & V1\\
\hline
 max efficiency & 63\% & 60\% & 55\% \\
 mean efficiency & 59\% & 56\% & 49\% \\
\hline\hline
\end{tabular} 
\end{center} 
\caption{The efficiency with which we can detect the injected supernova
core collapse waveform DFM\_A2B4G1 at a false alarm rate of 0.1 Hz.  The
upper line gives the maximum efficiency obtained by one of the search
algorithms for each detector.  The lower line gives the average
efficiency of the seven search methods used.}  
\label{tab:burst_single_eff} 
\end{table}

\subsection{Multi-interferometer analysis}

A true gravitational wave event will produce a signal in all detectors,
the amplitude of which will depend upon the location and polarization of
the source relative to the detector.  Furthermore, the time at which the
signal occurs in different detectors must differ by less than the light
travel time between the sites.  In contrast, false alarms caused by
noise will typically not occur simultaneously in several detectors.  By
requiring time coincidence between several sites, we should be able to
greatly reduce the number of false alarms.  However, we will also lose
some gravitational wave signals which are poorly aligned, and
consequently not detectable above the noise, in one or more of the
detectors.  The challenge is then to obtain the best possible efficiency
at a given false alarm rate.  There are two obvious coincidence options
--- to require a coincident signal in all three detectors, or to require
coincidence in only two of them.  Here, we will examine which of these
gives the better efficiency.

First, we can examine triple coincidences --- events which are seen in
all three of the Hanford, Livingston and Virgo detectors.  The single
interferometer false alarm rates and time coincidence windows lead to a
triple coincidence false alarm rate of around $1 \mu \mathrm{Hz}$.  At
this false alarm rate, the efficiency of the best performing algorithm
is 19\% while the average is 12\%.  At first sight, the triple
coincidence efficiency seems lower than expected.  However, consulting
Figure \ref{fig:dailyvariation} it is clear that there are significant
amounts of time when one of the three detectors is poorly aligned for
events from the galactic center and hence fairly insensitive to them.
So, there will be many events detected in two of the three detectors
which are not detected in the third.  This argues that we should also
look at the two detector results.  In order for this to be a fair
comparison, we perform a double coincident analysis for each pair of
detectors, with the same false alarm rate of $1 \mu \mathrm{Hz}$.  The
results are summarized in Table \ref{tab:burst_coinc_eff}.

\begin{table}[ht]
\begin{center}
\begin{tabular}{c c c c c c}
\hline \hline
 & HLV & HL & HV & LV & HL $\cup$ HV $\cup$ LV \\
\hline
 max efficiency & 19\% & 41\% & 22\% & 22\% & 60\% \\
 mean efficiency & 12\% & 31\% & 13\% & 15\% & 41\% \\
\hline\hline
\end{tabular} 
\end{center}
\caption{The efficiency with which we can detect the injections with
different combinations of detectors at a false alarm rate of $1 \mu
\mathrm{Hz}$.  The first column gives the triple coincidence efficiency.  The
next three give the efficiency of the various pairs of detectors.
Finally, we give the efficiency when we require an event to be detected
in two of the three detectors.  The false alarm rate in this case is
slightly higher at $\sim 3 \mu \mathrm{Hz}$.}
\label{tab:burst_coinc_eff} 
\end{table}

The two detector efficiencies are higher than the triple coincidence
efficiency, at the chosen false alarm rate.  Additionally, the
efficiency of the two LIGO detectors is higher than for one LIGO
detector with Virgo.  This is not surprising given the similar
orientations of the LIGO detectors.  Alternatively, we can combine the
two detector results to obtain an efficiency of 60\% for a search which
requires an event to be detected in at least two of the three detectors.
Note that the false alarm rate of this search may be somewhat higher
than the others ($\sim 3 \mu \mathrm{Hz}$ as compared to $1 \mu
\mathrm{Hz}$).  However, we expect that reducing the false alarm rate to
$1 \mu \mathrm{Hz}$ would have little to no effect on the efficiency.
From this preliminary study, we conclude that the best efficiency can be
obtained by requiring an event to be observed in two of the three
detectors.  This gives us a 60\% efficiency to the injected population,
in comparison to 41\% for the best combination of two detectors (H1-L1)
and a 19\% triple coincidence detection efficiency.  Thus, for this
population, a search using a network consisting of Virgo and the LIGO
detectors yields a 50\% greater efficiency than a LIGO only search.

Finally, it should be noted that the coincidence analysis described
above is only the first step towards a network analysis.  The optimal
network analysis would involve a coherent analysis of the data from all
detectors.  The application of various coherent methods to this
simulated data set, as well as a more detailed description of the
coincidence analysis will be presented in a future paper
\cite{ref:lv_burst_coh}.

\subsection{Directional Reconstruction}

\begin{figure}[h]
\includegraphics[width=18pc]{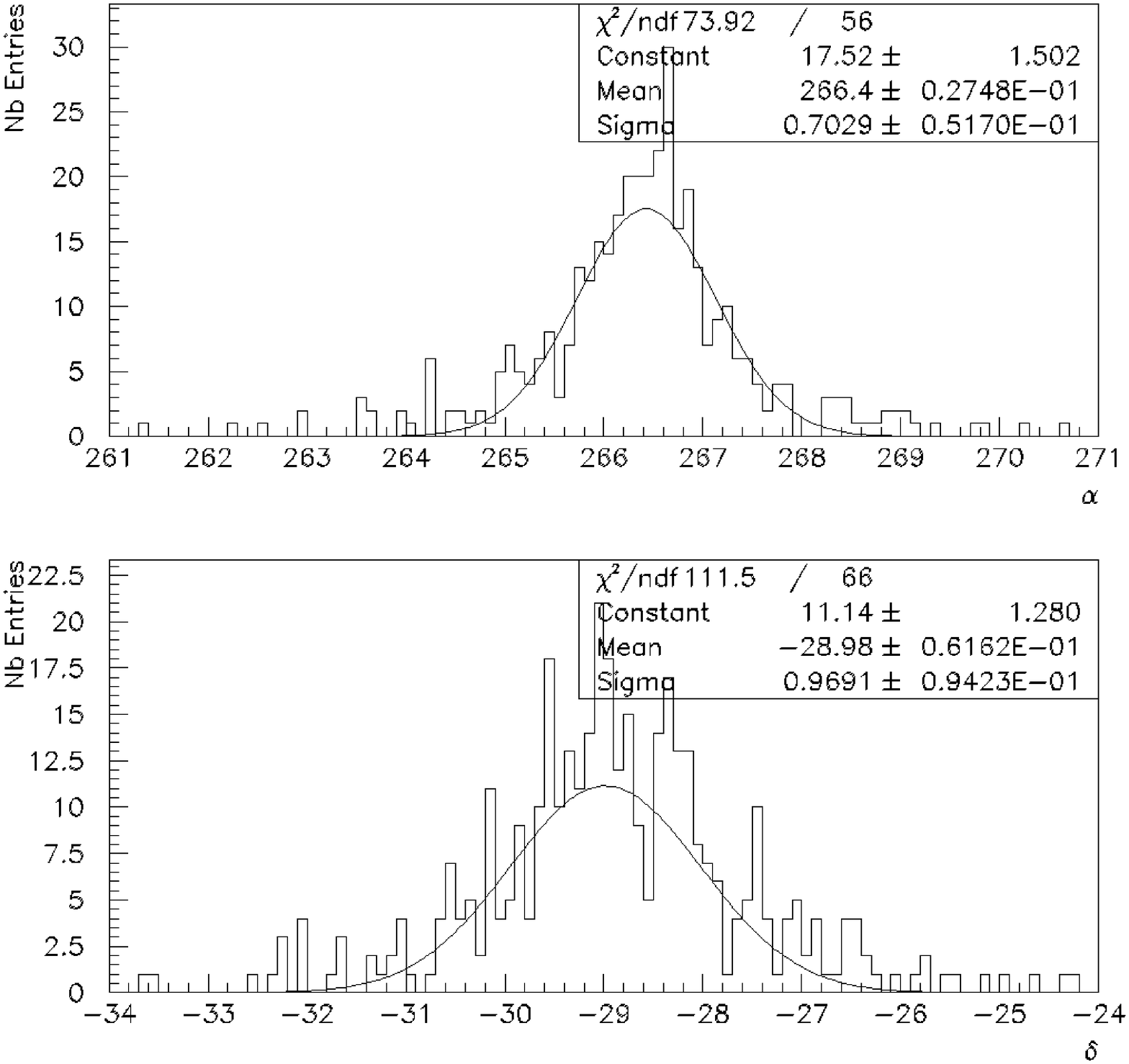}
\begin{minipage}[b]{18pc}
\caption{Histograms of the reconstructed sky location, right ascension
$\alpha$ and declination $\delta$, for Gaussian 1 ms injections detected
by the Peak Correlator.  The waveforms were injected from the direction
of the galactic center ($\alpha = 266.4^{\circ}$, $\delta =
-28.98^{\circ}$).}
\label{fig:burst_location}
\end{minipage}
\end{figure}

When an event is observed in three detectors, we can determine the sky
location using the timing information alone.%
\footnote{Timing information alone actually gives two sky positions.
The second location is the reflection of the true location in the plane
formed by the three detectors.  Here, we simply use the location closest
to the injected sky position.}
The accuracy with which we can determine the sky location is dependent
upon the resolution with which the arrival time of the signal can be
determined.  The arrival time of the signal is defined as the time of
the maximal amplitude.  For the 1ms Gaussian signals, analyzed with the
Peak Correlator, the arrival time can be determined to within 0.3~ms on
average (the time accuracy obtained by the Peak Correlator on Gaussian
signals can be parametrized as: $\sigma_{PC} = 1.43 10^{-4}
\frac{10}{SNR} \mathrm{ms}$).  Taking into account the observed SNR and
corresponding timing accuracy at each of the sites, we use a $\chi^{2}$
minimization technique to determine the sky location
\cite{ref:sky_location}.  In Figure \ref{fig:burst_location} we show the
accuracy with which the sky location can be reconstructed for a 1ms
Gaussian signal using the Peak Correlator.  The sky location is
determined quite accurately --- the average reconstructed value agrees
with the injected value and there is a one degree standard deviation in
both right ascension $\alpha$ and declination $\delta$.  For other
waveforms, particularly if they are not linearly polarized, our ability
to determine the arrival time and hence reconstruct the sky location may
vary considerably.

\section{Inspiral}

Gravitational waves from inspiralling binaries of neutron stars and/or
black holes are one of the most promising sources for the LIGO and Virgo
detectors.  Both the LIGO and Virgo collaborations have implemented
inspiral search pipelines, and a first comparison was made in
\cite{ref:lv_inspiral}.  In this section, as with the previous
discussion of burst searches, we will focus on the benefits of a
multi-detector and multi-site coincidence analysis.  We begin with a
description of the inspiral waveforms used in this analysis.  Following
a brief discussion of the single instrument results, we move on to
describe the coincidence analysis and directional reconstruction.  In
this study, we restrict attention to binary neutron star signals.

\subsection{Injections}

We can summarize the sensitivity of a detector to inspiral signals from
binary neutron star systems with a single number: the observable
effective distance, or range.  This is defined as the distance at which
an inspiral of $1.4-1.4 M_{\odot}$ neutron stars, in the optimal
direction and orientation with respect to each detector would produce a
signal to noise ratio of 8.  The effective distance of a signal is
always greater than or equal to the actual distance and on average is
about 2.3 times as large as the actual distance.  The ratio of effective
to actual distance depends upon the location of the source relative to
the detector, as well as the orientation (polarization and inclination)
of the source.  

At design sensitivity the inspiral ranges of the initial LIGO and Virgo
detectors are between 30 and 35 Mpc.  Consequently, the Virgo cluster,
at a distance of 16 Mpc, provides the largest concentration of galaxies
containing potential inspiral signals for the first generation of
gravitational wave interferometers.  In order to examine the benefits of
a joint network analysis, we inject inspiral signals from the the M87
galaxy in the Virgo cluster.  In addition, we add simulated signals from
a somewhat closer galaxy, namely NGC 6744 at 10 Mpc.  We inject a total
of 144 simulated events into the 24 hours of simulated data, with
approximately half of the events coming from each galaxy.  During the
course of the 24 hours of data, the location of the galaxies relative to
the detectors changes, thus allowing us to sample times when various
detectors are more and less sensitive to sources from these galaxies.
The component masses of the neutron stars in the binary are taken to be
between $1$ and $3 M_{\odot}$.  Furthermore, the inclincation,
polarization and coalescence phases are uniformly distributed among
their allowed values. 

\subsection{Single interferometer analysis}

Both the LIGO and Virgo collaborations  have implemented inspiral search
pipelines.  The LIGO pipeline has been used to search for binary
inspirals in the data taken during the first two LIGO science runs.
Details of the analysis pipeline and searches performed are available in
Refs.  \cite{ref:LIGO_S1_iul, ref:LIGO_S2_iul}.  The Virgo collaboration
has implemented two independent inspiral pipelines.  The first is a
standard flat search pipeline, ``Merlino'' \cite{ref:virgo_merlino},
while the second is a multiband templated analysis (MBTA).  In the
multiband approach, the templates are split for efficiency into high and
low frequency parts during the search \cite{ref:virgo_mbta}.  In
\cite{ref:lv_inspiral} we performed a first comparison of the LIGO and
MBTA pipelines.  

\begin{figure}
\includegraphics[width=18pc]{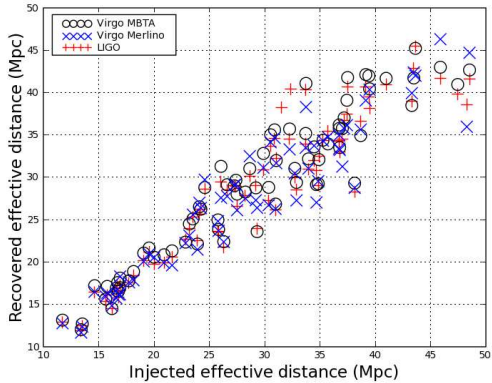}
\begin{minipage}[b]{18pc}

\caption{A comparison of injected and recovered effective distances for
the three inspiral pipelines.}

\label{fig:inspiral_eff_dist}
\end{minipage}
\end{figure}

The three pipelines were used to analyze the simulated data from all
three detectors. Since all the pipelines perform a matched filtering for
a specific waveform, we expect to obtain comparable results from all the
pipelines.  To verify this, we analyze the 24 hours of simulated data
plus injections with each of the three pipelines.  To make the results
directly comparable we use identical template bank generation parameters
and a signal to noise ratio threshold of $6$ for all searches.    As an
example, we look at the recovered effective distance of the simulated
events.  In Figure \ref{fig:inspiral_eff_dist} we show the injected and
recovered effective distances for the three pipelines.  The effective
distance is well recovered by all pipelines.  Indeed, for more distant
events, the difference between the values recovered by the three
pipelines is often less than the difference between the injected and
recovered distances.  This is because, at low signal to noise ratio, the
noise can have a significant effect on the recovered effective distance.
However, as all pipelines filter the same injections and the same noise,
we still expect good agreement between pipelines.  A more complete
comparison of the three pipelines will be presented in
\cite{ref:lv_inspiral_coh}.

For the inspiral search, we also examine the benefits of a network
search.  To simplify the presentation, we restrict to one pipeline,
namely the MBTA, for the remainder of this paper.  Since the results
obtained by the three pipelines are similar, our conclusions will not be
dependent on the pipeline used.  The MBTA single detector efficiencies
to the injected signals from the two source galaxies are given in Table
\ref{tab:inspiral_single_eff}.  The single instrument false alarm rates
at an SNR threshold of 6 are around $0.1 \, \mathrm{Hz}$.  The efficiency
of each of the three detectors is comparable for both galaxies.  As
expected, the efficiency to injections from NGC 6744 is larger as that
galaxy is closer than M87.  With the small number of injections
performed in this study, the diffferences between interferometers'
efficiencies are not significant.

\begin{table}[ht]
\begin{center}
\begin{tabular}{c c c c}
\hline \hline
 & H1 & L1 & V1\\
\hline
NGC 6744 efficiency & 72\% & 69\% & 68\% \\
M87 efficiency & 52\% & 57\% & 47\% \\
\hline\hline
\end{tabular} 
\end{center}
\caption{The efficiency of detecting inspiral injections from NGC 6744
(at a distance of 10 Mpc) and M87 in the Virgo cluster (at a distance of
16 Mpc) in the three detectors with a SNR threshold of 6.}
\label{tab:inspiral_single_eff} 
\end{table}

\subsection{Multi interferometer analysis}

The inspiral pipelines can accurately recover several parameters of the
injected waveforms, most notably the mass parameters, coalescence time
and effective distance (as shown in Fig. \ref{fig:inspiral_eff_dist}).
Since the effective distance can differ between detectors due to their
different orientations, we cannot use it when testing for coincidence.
We do require consistency of the coalescence time and mass between
signals in the detectors.  As with the burst search, this greatly
reduces the coincident false alarm rate.  However, due to the
coincidence test on mass, as well as time, it is difficult to estimate
the false alarm rate.  In the 24 hours of data searched, we find no
triple coincident, and only one double coincident, false alarm.

Next, we examine which combination of detectors gives the best detection
efficiency for our given injected population.  As with the burst search,
we consider the triple coincident search and various two detector
coincident searches.  The results are given in Table
\ref{tab:inspiral_coinc_eff}.  We use an SNR threshold of 6 in all
instruments.  This leads to the triple coincident search having a
substantially lower false alarm rate than the two detector searches.

\begin{table}[ht]
\begin{center}
\begin{tabular}{c c c c c c}
\hline \hline
 & HLV & HL & HV & LV & HL $\cup$ HV $\cup$ LV \\
\hline
 NGC 6744 efficiency & 48\% & 65\% & 54\% & 49\% & 72\% \\
 M87 efficiency & 24\% & 42\% & 32\% & 30\% & 56\% \\
\hline\hline
\end{tabular} 
\end{center} 
\caption{The efficiency of detecting inspiral injections from NGC 6744
(at a distance of 10 Mpc) and M87 in the Virgo cluster (at a distance of
16 Mpc) using different combinations of the LIGO and Virgo detectors and
an SNR threshold of 6 in all detectors.}  
\label{tab:inspiral_coinc_eff} 
\end{table}

The coincidence results clearly show the benefits of performing a search
including all three detectors.  The highest efficieny is obtained by
requiring a signal to be observed in any two of the three detectors.
For the closer NGC 6744 galaxy, the main advantage of adding the Virgo
detector to a LIGO only search is the good triple coincident efficiency.
Not only is the triple coincident false alarm rate very low, but also
with a trigger in three detectors we can reconstruct the sky location of
the source.

For signals from M87, the two detector LIGO efficiency is greater than
either the H1-V1 or L1-V1 efficiency.  This is expected due to the
similar orientations of the two LIGO detectors.  However, by including
Virgo and requiring a coincident trigger in two of the three detectors,
we do obtain a 25\% increase in efficiency.  The M87 galaxy is in the
Virgo cluster, which contains a significant fraction of potential binary
neutron star inspiral sources for the initial interferometric detectors.
A 25\% increase in efficiency to these sources significantly increases
the chance of making a detection.

\subsection{Directional Reconstruction}

\begin{figure}
\includegraphics[height=15pc]{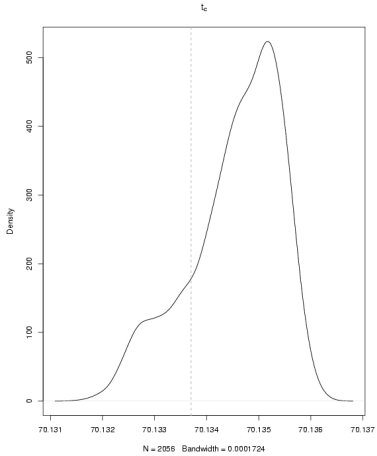}
\begin{minipage}[b]{18pc}
\caption{The posterior probability distribution for the end time of an
inspiral injection.  The dashed line shows the injected value.  The
width of the distribution, and hence the ability with which we can
determine the end time, is $\sim 5$ ms.}
\label{fig:inspiral_end_time}
\end{minipage}
\end{figure}

\begin{figure}
\includegraphics[width=18pc]{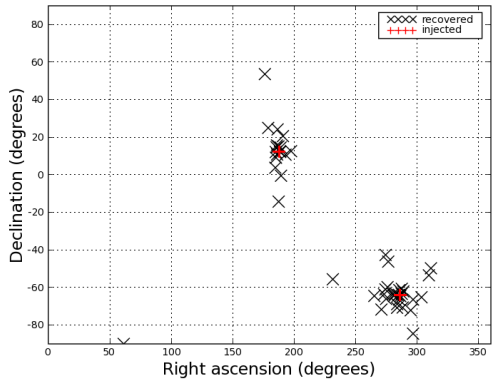}
\begin{minipage}[b]{18pc}
\caption{The recovered and injected sky locations of the inspiral
injections seen in all three detectors.  For reference, the galaxy NGC
is located at $\alpha = 286^{\circ}$, $\delta = -64^{\circ}$ and M87 is
located at $\alpha = 188^{\circ}$, $\delta = 12^{\circ}$.}      
\label{fig:inspiral_direction}
\end{minipage}
\end{figure}

In an inspiral search, the waveform can be parametrized by several
variables, among them the coalescence time, location, orientation and
mass parameters of the binary system.  It is well known that the
reconstructed values of these parameters are not independent.  For
example, a higher mass binary inspiral will traverse the sensitive band
of the detectors more rapidly than one of lower mass. Thus, the
reconstructed coalescence time and masses of the system will be
correlated.  These correlations make it difficult to determine the
coalescence time with good accuracy.  To illustrate this, we use a
Markov Chain Monte Carlo analysis \cite{ref:mcmc} to obtain the
posterior probability distribution of the various parameters.  In Figure
\ref{fig:inspiral_end_time} we show the distribution of the coalescence
time for one of the injections.  The width of this distribution is $\sim
5$ ms.  Due to the uncertainty in the coalescence time of the signal we
obtain a similar uncertainty in the reconstructed sky location.  Figure
\ref{fig:inspiral_direction} shows the accuracy with which we can
determine the sky location based on our coincidence search.  Using this
coincidence search, we cannot reconstruct the sky location with
sufficient accuracy to determine the galaxy containing the binary.

The coincidence search described above is only the first stage of a
network analysis.  The complete analysis would involve a coherent search
as a follow-up on any interesting candidates obtained from the
coincidence stage \cite{ref:coherent}.  In a coherent search, the data
from each detector is match filtered against a template with identical
mass parameters.  By using the same template to filter all the data, the
directional reconstruction can be separated from uncertainties in the
template mass parameters.  Therefore, it is likely that this coherent
step would improve our ability to recover the sky location of sources.
Details of the coherent search will be included in a future publication
\cite{ref:lv_inspiral_coh}.

\section{Discussion}

We have analyzed 24 hours of simulated data for the H1, L1 and V1
detectors.  By adding simulated burst and inspiral signals, we have
examined the benefits of performing a joint LIGO--Virgo coincidence
search for these sources.  We find two benefits of a LIGO--Virgo joint
search over a LIGO only or Virgo only search.  First, use of three
detectors substantially increases the efficiency to burst and inspiral
signals, at a fixed false alarm rate.  This increase is best realized by
requiring a signal to be observed in at least two of the three
detectors.  In addition, we can reconstruct the sky location of those
signals which are observed in all three detectors.  The accuracy with
which the direction can be determined is dependent upon the timing
accuracy of the search, with a 0.3 ms timing error leading to
approximately a $1^{\circ}$ uncertainty in the sky location.  The
results presented here show that there is significant benefit to
performing a joint coincidence search of LIGO and Virgo data.  


\section*{Acknowledgments}

LIGO Laboratory and the LIGO Scientific Collaboration gratefully aknowledge 
the support of the United States National Science Foundation for the 
construction and operation of the LIGO Laboratory and for the support of 
this research.



\section*{References} 

\begin{thebibliography}{99} 

\bibitem{ref:ligo}B. Abbott et al. (The LIGO Scientific Collaboration),
Nucl.  Instrum. methods Phys. Res. A \textbf{517}, 154 (2004).  \\

\bibitem{ref:virgo} F. Acernese et al. (The Virgo Collaboration), Class.
Quantum Grav. \textbf{21}, S385 (2004).

\bibitem{ref:geo}B. Willke et al., Class. Quantum Grav. \textbf{21},
S417 (2004).

\bibitem{ref:tama} R. Takahashi and the TAMA Collaboration, Class.
Quantum Grav. \textbf{21}, S403 (2004), M.\ Ando et al., Phys. Lett.,
\textbf{ 86}, 3950 (2001). 

\bibitem{ref:lt_burst} B.~Abbott et. al. (the LIGO Scientific
Collaboration) and T.~Akutsu et. al. (the TAMA Collaboration), ''Upper
limits from the LIGO and TAMA detectors on the rate of
gravitational-wave bursts,'' arXiv:gr-qc/0507081.

\bibitem{ref:lt_inspiral} S.~Fairhurst and H.~Takahashi (for the LIGO
Scientific Collaboration and the TAMA Collaboration), Class.
Quantum Grav. \textbf{22} S1109-S1118 (2005).

\bibitem{ref:lv_burst} F.~Beauville et al. (Joint LIGO/Virgo working
group), Class.  Quantum Grav. \textbf{22} S1293-S1302.

\bibitem{ref:lv_inspiral} F.~Beauville et al. (Joint LIGO/Virgo working
group), Class.  Quantum Grav. \textbf{22} S1149-S1158 (2005).

\bibitem{ref:dfm} H. Dimmelmeier, J.A. Font, E. Mueller, Astron.
Astrophys. 393, 523-542 (2002)

\bibitem{ref:lv_burst_coh} Joint LIGO/Virgo working group, ``Strategy
for performing an optimal joint LIGO--Virgo burst analysis'', in
preparation.

\bibitem{ref:sky_location} F. Cavalier et al. paper in preparation.

\bibitem{ref:LIGO_S1_iul} B. Abbott et al., (The LIGO Scientific
Collaboration), Phys. Rev. D \textbf{69}, 122001 (2004) .

\bibitem{ref:LIGO_S2_iul} B. Abbott et al., (The LIGO Scientific
Collaboration), ``Search for gravitational waves from galactic and
extra--galactic binary neutron stars", gr-qc/0505041.

\bibitem{ref:virgo_merlino} P. Amico, L. Bosi, C. Cattuto, L.
Gammaitoni, F. Marchesoni, M. Punturo, F. Travasso, H. Vocca, Comp.
Phys. Comm. \textbf{153}, 179 (2003).

\bibitem{ref:virgo_mbta} F. Marion et al., (The Virgo Collaboration),
Proceedings of the Rencontres de Moriond 2003, Gravitational Waves and
Experimental Gravity (2004).

\bibitem{ref:lv_inspiral_coh} Joint LIGO/Virgo working group,
``Strategy for performing an optimal joint LIGO--Virgo inspiral
analysis'', in preparation.

\bibitem{ref:mcmc} N. Christensen, A. Libson, and R. Meyer, Class.
Quantum Grav. \textbf{21} 317-330 (2004).

\bibitem{ref:coherent} A. Pai, S. Dhurandhar, and S. Bose
Phys. Rev. D \textbf{64}, 042004 (2001)

\end{thebibliography}
\end{document}